\documentclass[11pt]{article}
\usepackage{amsmath}
\usepackage{amsfonts}
\newcounter{mnotecount}[section]

\usepackage{epsf}
\usepackage{psfrag}
\usepackage{graphics}
\usepackage{mathrsfs} 

\renewcommand{\themnotecount}{\thesection.\arabic{mnotecount}}

\newcommand{\mnote}[1]
{\protect{\stepcounter{mnotecount}}$^{\mbox{\footnotesize
$
\bullet$\themnotecount}}$ \marginpar{
\raggedright\tiny\em
$\!\!\!\!\!\!\,\bullet$\themnotecount: #1} }

\def\d{{\mathrm{d}}}
\def\eqq{\stackrel{.}{=}}
\def\be{\begin{equation}}
\def\ee{\end{equation}}
\def\bea{\begin{eqnarray}}
\def\eea{\end{eqnarray}}
\def\scri{{\cal{I}}}
\def\r{\rho}
\def\t{\tau}

\def\eqq{\stackrel{\Omega}{=}}
\def\scri{{\mathscr{I}}}

\newtheorem{Teo}{Theorem}[section]
\newtheorem{Prop}[Teo]{Proposition}

\newtheorem{Def}[Teo]{Definition}
\newtheorem{Rem}[Teo]{Remark}

\begin{document}
\title{Formation of Higher-dimensional Topological Black Holes}
\author{Filipe C. Mena$^{1,2}$, Jos\'e Nat\'ario$^{2}$ and Paul Tod$^{3}$\\
{\small $^1$ Departamento de Matem\'atica,
Universidade do Minho,
4710-057 Braga, Portugal}\\ 
{\small $^2$ Departamento de Matem\'atica, Instituto Superior T\'ecnico, 1049-001 Lisboa, Portugal}\\
{\small $^3$ Mathematical Institute,
University of Oxford,
St Giles' 24-29, Oxford OX1 3LB, U.K.}}
\date{}
\maketitle
\begin{abstract}
We study higher dimensional gravitational collapse to topological black holes in two steps. 

Firstly, we construct some $(n+2)$-dimensional collapsing space-times, which include generalised Lema\^{\i}tre-Tolman-Bondi-like solutions, and we prove that these can be matched to static $\Lambda$-vacuum exterior space-times. We then investigate the global properties of the matched solutions which, besides black holes, may include the existence of naked singularities and wormholes.

Secondly, we consider as interiors classes of 5-dimensional collapsing solutions built on Riemannian Bianchi IX spatial metrics matched to radiating exteriors given by the Bizo\'{n}-Chmaj-Schmidt metric. 
In some cases, the data at the boundary for the exterior can be chosen to be close to the data for the Schwarzschild solution.  
\end{abstract}   
\section{Introduction}
Black holes in higher dimensions play an important role in theoretical physics, particularly in string theory. Although there has been work on both mathematical and physical aspects of higher dimensional topological black holes (see e.g. \cite{Birmingham, Gibbons-Hartnoll}), little has been done concerning the existence and stability of dynamical processes involved in their formation. This problem might be tackled by constructing appropriate matched spacetimes which settle through gravitational collapse to a topological black hole solution. Past approaches to this problem in 4-dimensions include the works of \cite{Lemos, Smith-Mann} for the collapse of Friedman-Lema\^\i tre-Robertson-Walker (FLRW) fluids and, more recently, \cite{MNT07} for the collapse of inhomogeneous and anisotropic fluids. 

In this paper, we consider this problem in higher dimensions. We start in section 2 by considering a family of solutions to the $\Lambda$-vacuum Einstein equations in $n+2$ dimensions which contains black hole solutions. We find some possible interior collapsing solutions with dust as source and study the corresponding matching problem. 
%

The metric exterior builds the $(n+2)$-dimensional space-time $M$ from a Riemannian $n$-dimensional Einstein manifold $N$. Black holes with this metric are difficult to integrate into the usual intuition of a black hole as a simple object formed in collapse and this is what the work in section 2 explores. The space-times are weakly-asymptotically-simple but not asymptotically-flat (or dS or AdS)\footnote{Since the metric on the sections of $\scri$ need not be a metric of constant curvature.}, which cannot happen in 4-dimensions. We shall seek to fill them in with dust solutions, so that they are formed by collapse, and we find large classes of examples which we present in section 2. 

When the Einstein manifold $N$ is not cobordant to a point (e.g.~$\mathbb{CP}^2$) the solutions we find cannot have a regular origin, though they can be regular with space-time wormholes or a `cusp' at the origin. When there is a singularity at the origin, it may or may not be visible from infinity.

All these filled-in solutions have static exteriors. As a step in the direction of constructing a dust collapse with a radiating exterior, we go on in section 3 to consider the 5-dimensional Bizo\'{n}-Chmaj-Schmidt (BCS) exterior \cite{BCS}, which has a deformed 3-sphere as the metric of constant $(t,r)$. The solution is known \cite{DafHol, Hol} to settle down via radiation to the round 3-sphere and the 5-dimensional Schwarzschild solution. We show how this exterior can be matched, at least in the neighbourhood of the matching surface, to one of a range of collapsing dust interiors. We also show that, for some of these interiors, the data at the matching surface can be chosen to be close to the data for the 5-dimensional Schwarzschild exterior. Since this solution is known to be stable \cite{DafHol}, it is reasonable to expect that the exterior will settle down to the Schwarzschild solution.

\section{Collapse to Black Holes without Gravitational Wave Emission} 

\subsection{A family of higher-dimensional black holes}
\label{sHDBH}
We start by stating our conventions.

\begin{Def}
An $(n+2)$-dimensional Lorentzian manifold $(M,g_{ab})$ is said to be a solution of the {\em Einstein equations with cosmological constant $\Lambda$ and energy-momentum tensor $T_{ab}$} if its Ricci tensor $R_{ab}$ satisfies
\[
R_{ab} = \Lambda g_{ab} + \kappa (T_{ab} - \frac1n T g_{ab})
\]
where $\kappa$ is a constant and $T=T^a_{\,\,\,a}$.
\end{Def}

\begin{Rem}
The Einstein equations can also be written as
\[
R_{ab} - \frac12 R g_{ab} + \frac{n\Lambda}2 g_{ab} = \kappa T_{ab},
\]
where $R=R^a_{\,\,\,a}$ is the Ricci scalar.
\end{Rem}
We wish to consider the family of higher-dimensional black holes given in the following proposition\footnote{This is a small generalisation of the metrics in \cite{gis}, clearly known to \cite{Gibbons-Hartnoll}.}.

\begin{Prop}
Let $(N,\d\sigma^2)$ be an $n$-dimensional Riemannian Einstein manifold with Ricci scalar $n\lambda$, and let
\be\label{v1}
V(r)=\frac{\lambda}{n-1}-\frac{2m}{r^{n-1}}-\frac{\Lambda r^2}{n+1},
\ee
where $m$ and $\Lambda$ are constants. If $I \subset \mathbb{R}$ is an open interval where $V$ is well defined and does not vanish then the $(n+2)$-dimensional Lorentzian manifold $(M,ds^2)$ given by $M=\mathbb{R} \times I \times N$ and
\be\label{m1}
\d s^2=-V(r)\d t^2+(V(r))^{-1}\d r^2+r^2\d \sigma^2,
\ee
is a solution of the vacuum Einstein equations with cosmological constant $\Lambda$.
\end{Prop}
\begin{Prop}
The metrics (\ref{m1}) are conformally-compactifiable at infinity.
\end{Prop}
\noindent{\bf{Proof:}} Using the null coordinate $u$ defined by
\be\label{m2}
\d u=\d t-\frac{\d r}{V}\ee
we can write the metric as
\be\label{m3}
\d s^2=-V\d u^2-2\d r\d u+r^2\d \sigma^2.
\ee
Setting $L=r^{-1}$ we have
\[{\d \hat s^2}:=L^2\d s^2=-L^2V\d u^2+2\d u\d L+\d\sigma^2.\]
Now clearly $\scri$ is at $L=0$, so that $\scri \sim\mathbb{R}\times N$. Note that $L^2V\sim-\Lambda/(n+1)$ as $L\rightarrow 0$ so that, as expected, $\scri$ is time-like, null or space-like according as $\Lambda<0,=0$ or $>0$. $\Box$
\\
\\
These metrics are weakly-asymptotically-simple, as we shall see below. However they are not, in general, asymptotically-flat (or dS or AdS) and we shall refer to them rather as \emph{asymptotically conical} (following \cite{Gibbons-Hartnoll}). 
\medskip

\begin{Prop}
The metrics (\ref{m1}) are weakly-asymptotically-simple.
\end{Prop}
\noindent{\bf{Proof:}} To see this, we look at null geodesics. If $x^i,i=1,\ldots,n$ are coordinates on $N$ then a geodesic in $M$ is given by first choosing a geodesic in $N$, say parameterised by arc-length as $x^i(\sigma)$; then choose $(t(s),r(s))$ and $\sigma(s)$ to satisfy
\begin{eqnarray*}
\dot{t}&=&E/V(r)\\
\dot\sigma &=&J/r^2\\
\dot r^2&=&E^2-\frac{J^2V}{r^2}
\end{eqnarray*}
where $E$ and $J$ are constants of integration. Written like this, the geodesic equations are formally identical to the null geodesic equations for the (four-dimensional) Schwarzschild solution. Thus, the `radial' null geodesics with $J=0$ and $\dot{r}=\pm E$ can be extended through the zeroes of $V$, when these exist, in the usual way by defining
\[\d u=\d t-\d r/V,\;\;\d v=\d t+\d r/V.\]
One may obtain complete extensions, parallelling the analogous cases in 4-dimensions. $\Box$

\medskip

\begin{Rem} \hspace{1cm}

\begin{itemize}
\item From the geodesic equations in the preceding proof one can see that for $m>0$ and $\lambda>0$ there are null geodesics at a fixed value of $r$ satisfying
\[r^{n-1}=\frac{m(n^2-1)}{\lambda},\]
though these won't be closed unless the corresponding geodesic on $N$ is closed.
\item With $\Lambda=0$, $\lambda >0$ and $m>0$, $V$ has a single zero, corresponding to an event horizon, and asymptotes to a positive constant at large $r$. This is a black-hole solution, which can be thought of as generalising the Schwarzschild metric. The (degenerate) metric on the horizon is $\d\sigma^2$, which is also the conformal metric on $\scri$.
\item With $\Lambda>0$, $m>0$ and large enough positive $\lambda$, $V(r)$ is positive between two zeroes, corresponding to a black-hole event horizon and a cosmological event horizon. The solution generalises the asymptotically-dS Kottler solution. 
\item With $\Lambda<0$ and $m>0$, $V$ again has a single zero, corresponding to an event horizon, and the solution generalises the asymptotically-AdS Kottler solution.
\item  The solutions in the previous class with $\lambda\leq 0$ may have no global symmetries except the staticity Killing vector. This is because compact, negative scalar curvature Einstein manifolds have no global symmetries, nor does, for example, the Ricci-flat metric on $K3$ (an example with  $\lambda=0$ and $n=4$).
\end{itemize}
\end{Rem}
\subsection{Possible interiors}
\label{sFRWLI}

\subsubsection{Some Einstein metrics}
\label{ssSEM}
We construct some Einstein $(n+1)$-metrics in the familiar way as cones on Einstein $n$-metrics.

\begin{Prop} If $I \subset \mathbb{R}$ is an open interval, $f:I \to \mathbb{R}$ is a positive smooth function and $(N, \d\sigma^2)$ is as before, then the Riemannian metric defined on the $(n+1)$-dimensional manifold $I \times N$ by
\be\label{met1}
\d \r^2+f(\r)^2\d\sigma^2
\ee
is Einstein with Ricci scalar $k(n+1)$ precisely in the following cases:
\begin{enumerate}
\item If $k>0$ then
\[k=\nu^2n,\; \lambda=\nu^2(n-1),\; f=\sin (\nu \r)\]
for some $\nu > 0$.
\item If $k=0$ then
\[\lambda=n-1,\;f=\r.\]
\item If $k<0$ then
\[k=-\nu^2n,\;\lambda=\nu^2(n-1),\;f=\sinh(\nu\r),\]
or
\[k=-\nu^2n,\;\lambda=0,\;f=e^{\pm\nu\rho},\]
or
\[k=-\nu^2n,\;\lambda=-\nu^2(n-1),\;f=\cosh(\nu \r),\]
for some $\nu > 0$, according as $\lambda>0$, $\lambda=0$ or $\lambda<0$.
\end{enumerate}
\end{Prop}
These metrics typically have singularities at the origin $\rho=0$. 

\begin{Prop}
The Kretschmann scalar $K$ of the $(n+1)$-metric above (that is the trace of the square of the Riemann tensor) is related to the square $C^2$ of the Weyl tensor of the base $n$-metric by
\be\label{c1}K=\frac{C^2}{f^4}+\mathrm{const}.\ee
\end{Prop}
Therefore the $(n+1)$-metric is singular anywhere $f$ vanishes, unless the $n$-metric is conformally-flat (like an $n$-sphere with the standard metric\footnote{Notice that there exist Einstein metrics on certain spheres which are not conformally flat \cite{Bohm}, which could be used here and elsewhere in this article.}). This can be avoided in case 3 with $f=e^{\pm\nu\rho}$ when the metric has an internal infinity (or a cusp) or $f=\cosh(\nu \r)$ when the metric has a minimal surface and a second asymptotic region, which will correspond to a space-time wormhole. Otherwise, if $f$ has a zero at which the metric is singular, we shall need to check whether this singularity is visible from infinity in the resulting space-time.
\subsubsection{Some FLRW-like metrics}
\label{ssFRW}
The previous subsection suggests a family of $(n+2)$-dimensional FLRW-like metrics.
\begin{Prop}
The $(n+2)$-dimensional Lorentzian metric
\be\label{met2}
\d s^2=-\d \t^2+R^2(\t)(\d \r^2+f^2(\r)\d\sigma^2).
\ee
is a solution of the Einstein equations with cosmological constant $\Lambda$ and energy-momentum tensor $T_{ab}=\mu \, u_au_b$, corresponding to a dust fluid with density $\mu$ and velocity $u_a\d x^a=\d \t$, if and only if $R(\t)$ and $\mu(\t)$ satisfy the conservation equation
\be\label{e1}
\mu R^{n+1}=\mu_0,
\ee
for constant $\mu_0$, and the Friedman-like equation
\be\label{e2}
\frac{\dot{R}^2}{R^2}+\frac{k}{nR^2}=\frac{2\kappa\mu}{n(n+1)}+\frac{\Lambda}{n+1}.
\ee
\end{Prop}
\begin{Rem}
There are other solutions of this type which we shall exploit below in section 3, namely
\be\label{met22}
\d s^2=-\d \t^2+R^2(\t)h_{ij}\d x^i\d x^j,
\ee
where $h_{ij}\d x^i\d x^j$ is chosen to be any Einstein $(n+1)$-metric with Ricci scalar $k(n+1)$. Explicitly we shall take the Einstein metric to be one of Eguchi-Hansen \cite{EH}, $k$-Eguchi-Hansen \cite{Pederson} or $k$-Taub-NUT \cite{bj}. FLRW-like dust cosmologies are again given by solutions of (\ref{e1})-(\ref{e2}).
\end{Rem}

\subsubsection{Some Lema\^itre-Tolman-Bondi-like solutions}
\label{ssTB}
We shall now introduce Lema\^itre-Tolman-Bondi-like (LTB-like) solutions, generalising those of \cite{gb}, related to the metrics of subsection \ref{ssFRW}.

\begin{Prop}
The $(n+2)$-dimensional Lorentzian metric
\be\label{m5}
\d s^2=-\d\t^2+A(\t,\r)^2\d \r^2+B(\t,\r)^2\d\sigma^2,
\ee
is a solution of the Einstein equations with cosmological constant $\Lambda$ and energy-momentum tensor $T_{ab}=\mu \, u_au_b$, corresponding to a dust with density $\mu$ and velocity $u_a\d x^a=\d \t$, if and only if $A(\t,\r)$, $B(\t,\r)$ and $\mu(\t,\r)$ satisfy
\bea
A&=&B'(1+w(\r)),\label{mm1}\\
\mu AB^{n}&=&M'(\r)(1+w(\r)),\label{mm2}
\eea
for some functions $w(\rho)$ and $M(\rho)$, and
\be\label{m4}\dot{B}^2B^{n-1}+\left(\frac{\lambda}{n-1}-\frac{1}{(1+w(\r))^2}-\frac{\Lambda}{n+1}B^2\right)B^{n-1}=\frac{2\kappa M(\r)}{n}\ee
(where dot and prime denote differentiation with respect to $\tau$ and $\rho$).
\end{Prop}
\begin{Rem}
This metric has three free functions of $\r$, namely $w(\r),M(\r)$ and $B(0,\r)$, one of which can be removed by coordinate freedom.
\end{Rem}
%
\subsection{Matching}
In this subsection, we seek to match an interior represented by the metric (\ref{met2}) to a static exterior represented by the metric (\ref{m1}) at a surface $\Omega$ ruled by radial time-like geodesics in (\ref{m1}) which is comoving, i.e. a surface of constant $\rho$ (say $\rho=\r_0$) in (\ref{met2}). We find that this can be done, subject to conditions found below.
\begin{Prop} \label{matching1}
The metric (\ref{m1}) can be matched to the FLRW-like metric (\ref{met2}) at $\rho=\rho_0$ provided that $f'(\rho_0)>0$ and 
$\displaystyle{m=\frac{\kappa\mu_0f(\r_0)^{n+1}}{n(n+1)}}$.
\end{Prop}
\noindent{\bf{Proof:}}
The interior metric on the matching surface $\Omega$ is
\[-\d\t^2+R(\t)^2f(\r_0)^2\d\sigma^2,\]
while the geodesic in the exterior, parameterised by proper time $\t$, has
\[\dot{t}=\frac{E}{V},\;~~\dot{r}^2=E^2-V,\]
and the exterior metric on $\Omega$ becomes
\[-\d\t^2+r(\t)^2\d\sigma^2.\]
Introducing $\eqq$ to mean equal at $\Omega$ we must then have
\be\label{mat1}r\eqq R(\t)f(\r_0).\ee
For the second fundamental form, the matching reduces to a calculation already done in \cite{MNT07}, and is
\be\label{mat2}\frac{V\dot{t}}{r}\eqq\frac{f'}{Rf},\ee
which, with (\ref{mat1}) and the geodesic equation, reduces to
\be\label{mat3}E=f'(\r_0).\ee
Since we need $E>0$, this constrains the matching to a region where $f'>0$. The other geodesic equation, with the dot of (\ref{mat1}), is
\[\dot{r}^2=E^2-V=\dot{R}^2f(\r_0)^2,\]
which, with (\ref{v1}), reduces to the Friedman equation (\ref{e2}) if we make the identifications
\be\label{c13}
m=\frac{\kappa\mu_0f(\r_0)^{n+1}}{n(n+1)},
\ee
and
\[E^2=\frac{\lambda}{n-1}-\frac{kf(\r_0)^2}{n}.\]
The first of these determines the mass $m$ of the exterior from the density and size of the interior. The second is an identity, as can be checked from the formulae in section \ref{ssSEM}. $\Box$

\medskip

\begin{Rem}
We shall use the term FLRW-Kottler space-times for these matched solutions. Since the matching requires $m>0$ in the exterior and $V>0$ at $\Omega$, we can have FLRW-Kottler space-times with any sign on $\Lambda$ for $\lambda>0$, but if $\lambda\leq 0$ then the matching requires $\Lambda<0$.
\end{Rem}

\begin{Prop} \label{matching2}
The metric (\ref{m1}) can be matched to the LTB-like metric (\ref{m5}) at $\rho=\rho_0$ provided that $1+w(\r_0)>0$ and 
$\displaystyle{m=\frac{\kappa}{n}M(\rho_0)}$.
\end{Prop}
\noindent{\bf{Proof:}}
The proof is analogous to the previous one, and we obtain
\begin{eqnarray*}
r&\eqq&B(\t,\rho_0),\\
E&=&(1+w(\r_0))^{-1},\\
m&=&\frac{\kappa}{n}M(\rho_0),
\end{eqnarray*}
in place of (\ref{mat1}), (\ref{mat3}), and (\ref{c13}) respectively. $\Box$

\begin{Rem}
We shall use the term LTB-Kottler space-times for these matched solutions.
\end{Rem}
%
%
\subsection{Global Properties}
%
We will now analyse in detail the global properties of the FLRW-Kottler spacetime in the three cases $\Lambda=0$, $\Lambda>0$ and $\Lambda<0$, and make some remarks about the global properties of the LTB-Kottler spacetime.

\begin{figure}[h!]
\begin{center}
\psfrag{I+}{$\mathscr{I^+}$}
\psfrag{I-}{$\mathscr{I^-}$}
\psfrag{(a)}{(a)}
\psfrag{(b)}{(b)}
\epsfxsize=.6\textwidth
\leavevmode
\epsfbox{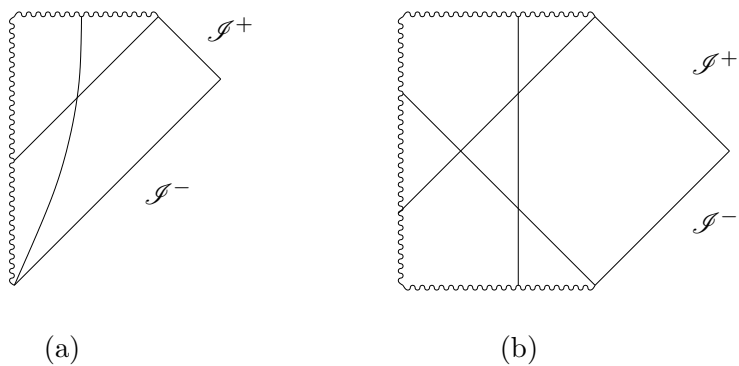}
\end{center}
\caption{Penrose diagram for $\Lambda = 0$ and (a) $k\leq 0$; (b) $k>0$, showing the matching surfaces and the horizons.}\label{Zero_Lambda}
\end{figure}

\begin{Prop}
If $\Lambda = 0$ (hence $\lambda > 0$) and $(N, \d \sigma^2)$ is not an $n$-sphere then the locally naked singularity of the FLRW-Kottler spacetime at $\r = 0$ is always visible from $\scri^+$ for $k \leq 0$, but can be hidden if $k>0$ and $n \geq 4$ (space-time dimension $n+2 \geq 6$).
\end{Prop}
\noindent{\bf{Proof:}} The first statement is clear from the Penrose diagram depicted in Figure~\ref{Zero_Lambda}. For the second statement we must compare the conformal lifetime of the FLRW universe
\[
\Delta T = 2 \int_0^{R_{max}} \frac{\d R}{R\dot{R}} = \frac{2\pi}{\nu(n-1)}
\]
with the supremum of the possible values of $\r_0$, which is $\frac{\pi}{2 \nu}$. For the singularity to be hidden it is necessary that the radial light ray emanating from $\r=0$ at the Big Bang is to the future of the future event horizon, and it is clear that in this case one will have $\rho_0 > \frac{\Delta T}{2}$. This is only possible if $\frac{\pi}{2 \nu}>\frac{\pi}{\nu(n-1)}$, i.e.~$n>3$. $\Box$

\begin{Rem}
A special role for space-time dimension $n+2 = 6$ in dust collapse was also found in \cite{gj}.
\end{Rem}

\begin{Prop}
If $\Lambda > 0$ (hence $\lambda > 0$) and $(N, \d \sigma^2)$ is not an $n$-sphere then the locally naked singularity of the FLRW-Kottler spacetime at $\r = 0$ can be always be hidden except if the FLRW universe is recollapsing (hence $k>0$) and $n < 4$.
\end{Prop}
\noindent{\bf{Proof:}} If the FLRW universe is recollapsing then one can show that its conformal lifetime is an increasing function of $\Lambda$, and approaches $\frac{2\pi}{\nu(n-1)}$ as $\Lambda \to 0$. Therefore the singularity can be hidden for sufficiently small $\Lambda$ if $n \geq 4$, but not if $n<4$. If the FLRW universe is not recollapsing then one can show that its conformal lifetime is a decreasing function of $\Lambda$ which approaches zero as $\Lambda \to + \infty$. Therefore the singularity can be hidden for sufficiently large $\Lambda$.

\begin{figure}[h!]
\begin{center}
\psfrag{I+}{$\mathscr{I^+}$}
\psfrag{I-}{$\mathscr{I^-}$}
\psfrag{(a)}{(a)}
\psfrag{(b)}{(b)}
\epsfxsize=.8\textwidth
\leavevmode
\epsfbox{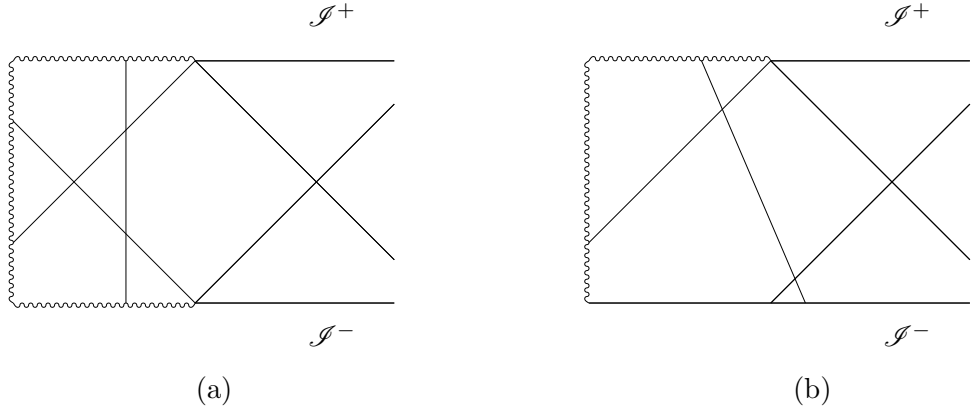}
\end{center}
\caption{Penrose diagram for $\Lambda > 0$ with the FLRW universe (a) recollapsing; (b) non-recollapsing, showing the matching surfaces and the horizons.}\label{Positive_Lambda}
\end{figure}

\begin{Prop}
For $\Lambda < 0$ the FLRW-Kottler spacetime satisfies the following:
\begin{enumerate}
\item If $\lambda > 0$ and $(N, \d \sigma^2)$ is not an $n$-sphere then the locally naked singularity of the FLRW-Kottler spacetime at $\r = 0$ can always be hidden.
\item If $\lambda = 0$ then the cusp singularity is not locally naked.
\item If $\lambda < 0$ then no causal curve can cross the wormhole from one $\scri$ to the other.
\end{enumerate}
\end{Prop}
\noindent{\bf{Proof:}} To prove the first statement one just has to check that the conformal lifetime of the FLRW universe goes to zero as $\Lambda \to - \infty$. Therefore one can always hide the singularity by taking $\Lambda$ small enough. The second statement follows from the fact that for $\lambda=0$ one must have $f(\r)=e^{\nu \r}$, and hence the cusp singularity is at $\r = -\infty$. To prove the third statement (which can be seen, essentially, as a corollary of a result of Galloway \cite{gal}) one notices that the future horizons hit the matching surfaces at marginally outer trapped surfaces. The set of all these surfaces forms the curve $\dot{R} + \nu \tanh (\nu \r) = 0$, which can be seen to be spacelike with the help of the Friedman-like equation \eqref{e2}. A similar argument shows that the past horizons are connected by the spacelike curve of marginally anti-trapped surfaces. The statement now follows from the observation that these two curves touch at $\dot{R}=\rho=0$. $\Box$

\begin{figure}[h!]
\begin{center}
\psfrag{I}{$\mathscr{I}$}
\psfrag{I1}{$\mathscr{I}_1$}
\psfrag{I2}{$\mathscr{I}_2$}
\psfrag{(a)}{(a)}
\psfrag{(b)}{(b)}
\psfrag{(c)}{(c)}
\epsfxsize=\textwidth
\leavevmode
\epsfbox{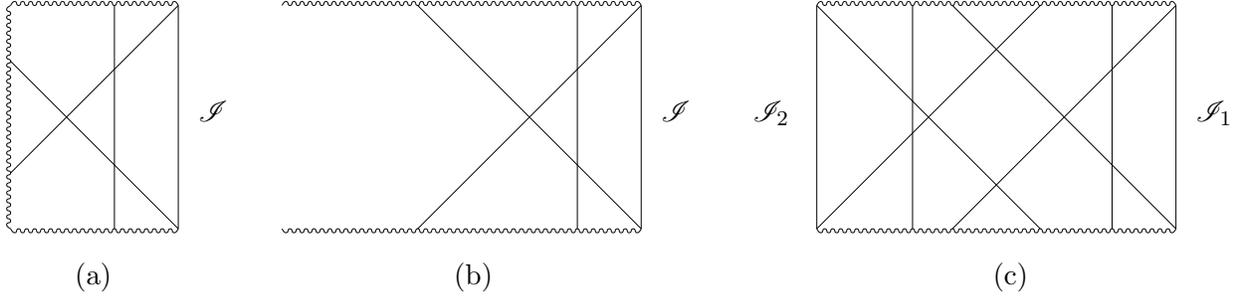}
\end{center}
\caption{Penrose diagram for $\Lambda < 0$ and (a) $\lambda > 0$; (b) $\lambda = 0$; (c) $\lambda < 0$, showing the matching surfaces and the horizons.}\label{Negative_Lambda}
\end{figure}

\begin{Rem}
The global properties of the LTB-Kottler spacetime obtained in Proposition \ref{matching2} are much more diverse. For instance, one can easily find examples of black hole formation with wormholes inside the matter with positive $\lambda$ and $\Lambda=0$ (similar results in 4-dimensions are in \cite{hell}). Indeed, take data
\[B(0,\rho)=a^2+\rho^2,\;\dot B(0,\rho)=0,\;A(0,\rho)=1.\]
Then $\t=0$ is a surface of time symmetry, the metric on $\t=0$ has a minimal surface at $\rho=0$, and $(1+w)^{-1}=2\rho$. Equation (\ref{m4}) becomes
\[\dot{B}^2=-\left(\frac{\lambda}{n-1}-4\rho^2\right)+\frac{2\kappa M(\r)}{n}B^{1-n}. \]
We restrict $\rho$ so that the first term is strictly negative,
\be\label{rr1}\rho^2<\frac{\lambda}{4(n-1)},\ee
so that $B$ necessarily expands from an initial zero, through a maximum at the moment of time symmetry to a final singularity. Note that $M(\r)$ is fixed by the condition $\dot B(0,\rho)=0$ to be
\[M(\rho)=\frac{n}{2\kappa}(a^2+\rho^2)^{n-1}\left( \frac{\lambda}{n-1}-4\rho^2\right).\]
From (\ref{mm2}) we calculate
\[\mu(0,\rho)=\frac{n}{2\kappa}\frac{(\lambda-4a^2-4n\rho^2)}{(a^2+\rho^2)^2}\]
and for this to be positive we must impose another condition on $\rho$:
\be\label{rr2}\rho^2<\frac{\lambda-4a^2}{4n}.\ee
If we can ensure that there is no shell-crossing and then match to the exterior at a value of $\rho$ satisfying the two restrictions (\ref{rr1}) and (\ref{rr2}) then we have formation of a black hole for $\lambda>0$ with $\Lambda=0$ and a wormhole.

To rule out shell-crossing, which would occur at a zero of $A$ in \eqref{m5}, we consider the $(\rho,\rho)$ component of the Einstein equations. This is
\[\frac{\ddot A}{A}+n\frac{\dot A\dot B}{AB}-\frac{n}{AB}\left(\frac{B'}{A}\right)'=\frac{\kappa\mu}{n},\]
so that
\[\frac{\partial}{\partial\t}\left(B^n\dot A\right)=2nB^{n-1}+\frac{\kappa M'}{2n\rho}.\]
The right-hand-side is positive so that $\dot A>0$ for $\t>0$ and so $A$ is never zero, i.e. there is no shell-crossing, for $\t\geq 0$. Since $A$ is an even function of $\t$ this shows that there is no shell-crossing at all for this example.
\end{Rem}

\section{Collapse to Black Hole with Gravitational Wave Emission}
The matchings in the previous section involved static exteriors. In this section, we shall consider gravitational collapse with a gravitational wave exterior, so that the exterior metrics will be time-dependent generalisations of those in (\ref{m1}). For simplicity, we confine our attention to one example, the Bizo\'n-Chmaj-Schmidt metric in $(4+1)$-dimensions \cite{BCS}, though a similar ansatz can be made in other dimensions and with other symmetries (see \cite{bcr}). We shall consider three different interiors with this exterior, built on Riemannian Bianchi type IX spatial metrics.
\subsection{The Exterior: Bizo\'n-Chmaj-Schmidt metric}
Consider the metric \cite{BCS}
\begin{equation}
\label{BCS-metric}
ds^{2+}=-Ae^{-2\delta}dt^2+A^{-1}dr^2+\frac{r^2}{4}e^{2B}(\sigma_1^2+\sigma_2^2)+\frac{r^2}{4}e^{-4B}\sigma_3^2
\end{equation}
where $A, \delta$ and $B$ are functions of $t$ and $r$. The one-forms $\sigma_i$ are left-invariant for the standard Lie group structure on $S^3$, satisfy the differential relations $d\sigma_i=\frac{1}{2}\epsilon_{ijk}\sigma_j\wedge \sigma_k$, and can be taken to be
\begin{eqnarray}
\label{sigmas}
\sigma_1 &= & \cos\psi d\theta+\sin\theta\sin\psi d\phi\nonumber\\
\sigma_2&=& \sin\psi d\theta -\sin\theta\cos\psi d\phi\\
\sigma_3&=& d\psi+\cos\theta d\phi\nonumber
\end{eqnarray}
where $\theta,\psi,\phi$ are Euler angles on $S^3$ with $0<\theta<\pi$, $0<\phi<2\pi$ and $0<\psi<4\pi $. The Schwarzschild limit of (\ref{BCS-metric}) is obtained by setting $B=0$. The space-time with $B\neq 0$ is interpreted as containing pure gravitational waves with radial symmetry \cite{BCS}. Note that there is a residual coordinate freedom
\be \label{delta}
t\rightarrow\hat t=f(t); \quad \quad \delta\rightarrow\hat \delta=\delta + \log\dot{f}
\ee
in the metric (\ref{BCS-metric}), which one can exploit to choose $\delta$ arbitrarily along a timelike curve.

The $(4+1)$-dimensional vacuum EFEs give
\begin{eqnarray}
\partial_r A& =& -\frac{2A}{r}+\frac{1}{3r} (8e^{-2B}-2e^{-8B})-2r(e^{2\delta}A^{-1}(\partial_t B)^2+A(\partial_r B)^2)\label{EFE-BCS1}\\
\partial_t A&=& -4rA(\partial_t B)(\partial_r B)\label{EFE-BCS2}\\
\partial_r\delta&=&-2r(e^{2\delta}A^{-2}(\partial_tB)^2+(\partial_rB)^2)\label{EFE-BCS3}
\end{eqnarray}
together with the quasi-linear wave equation for $B$
\be\label{ev4}
\partial_t(e^{\delta} A^{-1}r^3(\partial_t B))-\partial_r(e^{-\delta}Ar^3(\partial_r B))+\frac{4}{3}e^{-\delta}r(e^{-2B}-e^{-8B})=0.
\ee
In \cite{BCS} the authors solve this system by giving $B$ and $\partial_t B$ at $t=0$ with $A(0,0)=1$ and $\delta(t,0)=0$. We shall be interested in giving data $A$, $B$ and the normal derivative $\nabla_nB$ at the timelike boundary $\Omega$ of the collapsing interior, which is noncharacteristic for this system, with the gauge choice $\delta\eqq 0$. Uniqueness and local existence follow as standard. From \cite{Hol}, \cite{DafHol} one knows that the $5$-dimensional Schwarzschild metric is stable among the BCS solutions, so that if data close to that for Schwarzschild is given on an asymptotically-flat hypersurface then the solution will exist forever and stay close to the Schwarzschild solution. As we shall see, data on $\Omega$ can be chosen to be close to data for Schwarzschild. This is not sufficient to deduce that the solution exists forever and is aymptotically-flat in the exterior, but it makes it rather plausible.
\\\\
The matching surface is parameterised by
\[
\Omega^+=\{t(\tau),r(\tau)\},
\]
and the first fundamental form on $\Omega^+$ is
\[
ds^{2+}|_{\Omega^+}=-\d\tau^2+\frac{r^2}{4}e^{2B}(\sigma_1^2+\sigma_2^2)+\frac{r^2}{4}e^{-4B}\sigma_3^2.
\]
The normal vector to the matching surface is
\[
n^{a+}=\dot r\frac{e^\delta}{A}\partial_t+Ae^{-\delta}\dot t\partial_r,
\] 
where the dot denotes differentiation with respect to the parameter $\tau$.
The boundary as seen from the exterior is ruled by geodesics which obey
\begin{equation}
\label{geod}
Ae^{-2\delta}\dot t^2-A^{-1}\dot r^2=1.
\end{equation}
The second fundamental form on $\Omega^+$ reads
\begin{eqnarray}
K^+_{11}=K_{22}^+& =& \frac{r^2}{4}e^{2B}\nabla_nB+\frac{r}{4}e^{2B} A e^{-\delta}\dot t,\nonumber\\
K^+_{33} &=& -\frac{r^2e^{-4B}}{2}\nabla_nB+\frac{r}{4}e^{-4B} A e^{-\delta}\dot t.\nonumber
\end{eqnarray}
\subsection{The Interiors}
As interior metrics, we shall consider three classes of FLRW-like solutions based on Riemannian Bianchi-IX spatial metrics which are respectively the Eguchi-Hanson metric (with $R_{ij}=0$), the $k$-Eguchi-Hanson metric (with $R_{ij}=kg_{ij}$ excluding the case $k=0$) and the $k$-Taub-NUT metric (with $R_{ij}=kg_{ij}$, including $k=0$ as a particular case).
We summarize our results as follows:
\begin{Teo} In each case, the interior metric gives consistent data for the metric (\ref{BCS-metric}) at a comoving time-like hypersurface. Local existence of the radiating exterior in the neighbourhood of the matching surface is then guaranteed. In the case of Eguchi-Hanson and $k$-Taub-NUT with $k<0$, the data can be chosen to be close to the data for the Schwarzschild solution.

\end{Teo}

\subsubsection{The Eguchi-Hanson metric} 
Eguchi and Hanson found a class of self-dual solutions to the Euclidean Einstein equations with metric given by \cite{EH}
\begin{equation}
\label{EH-metric}
h_{EH}=\left(1-\frac{a^4}{\rho^4}\right)^{-1}d\rho^2+\frac{\rho^2}{4}(\sigma_1^2+\sigma_2^2)+\frac{\rho^2}{4}\left(1-\frac{a^4}{\rho^4}\right)\sigma^2_3
\end{equation}
with $\sigma_i$ given by (\ref{sigmas}) and $a$ is a real constant. The level sets of $\rho$ are topologically $S^3/\mathbb{Z}_2$, rather than $S^3$, but the corresponding quotient can also be taken on the metric \eqref{BCS-metric}.
\\\\
The FLRW-like metric built on this is
\[ds^{2-}=-\d \t^2+R^2(\t)h_{EH},\]
with the Einstein equations for a dust source reducing to
\be\label{ev6}\mu R^4=\mu_0,\;\dot{R}^2=\frac{\kappa\mu_0}{6R^2}.\ee
We shall match at $\rho=\rho_0$ so that $\Omega^-$ is parameterised by
\[
\Omega^-=\{\tau,\rho=\rho_0\}.
\]
The corresponding first fundamental form on $\Omega^-$ is
\[
ds^2|_{\Omega^-}= -\d\t^2+R^2(\tau)\left(\frac{\rho^2}{4}(1-\frac{a^4}{\rho^4})\sigma^2_3+\frac{\rho^2}{4}(\sigma_1^2+\sigma_2^2)\right),
\]
and the equality of the first fundamental forms then gives
\begin{eqnarray}
\label{EH-1}
r &\eqq& R\rho  e^{-B},\nonumber\\
e^{-6B} &\eqq&1-\frac{a^4}{\rho^4}.
\end{eqnarray}
The normal vector to the matching surface is
\[
n^{-}=\frac{1}{R}\left(1-\frac{a^4}{\rho^4}\right)^{\frac{1}{2}}\partial_\rho,
\]
and the associated non-zero components of the second fundamental form on $\Omega^-$ are
\begin{eqnarray}
K_{11}^-=K_{22}^-&\eqq& \frac{\rho R}{4}\left(1-\frac{a^4}{\rho^4}\right)^{\frac{1}{2}},\nonumber\\
K_{33}^- &\eqq&   \frac{R}{4}\left(\rho+\frac{a^4}{\rho^3}\right)\left(1-\frac{a^4}{\rho^4}\right)^{\frac{1}{2}}.\nonumber
\end{eqnarray}
The equality of the second fundamental forms gives
\begin{eqnarray}
\label{EH-2}
\nabla_{n}B&\eqq& -\frac{2a^4}{3R\rho^5}\left(1-\frac{a^4}{\rho^4}\right)^{-\frac{1}{2}}, \nonumber\\
Ae^{-\delta} \dot t &\eqq& e^{2B} \left(1-\frac{a^4}{3\rho^4}\right).
\end{eqnarray}
Then from (\ref{geod}), (\ref{ev6}) and (\ref{EH-1}) we calculate
\be\label{A1}
A\eqq e^{4B}\left(1-\frac{a^4}{3\rho^4}\right)^2-\frac{\kappa\mu_0}{6r^2}\rho^4e^{-4B}.
\ee
From the EFEs (\ref{EFE-BCS1})-(\ref{EFE-BCS3}) on $\Omega$ we get (using the matching conditions)
\[
\partial_t B\eqq -\nabla_n B\dot re^{-\delta},\quad \partial_r B\eqq\nabla_n B\dot t e^{-\delta},
\]
and then it is straightforward to check that the expression (\ref{A1}) for $A$ is consistent with $\dot A$ calculated from (\ref{EFE-BCS1}) and (\ref{EFE-BCS2}).

At this point, we have $B,\nabla_nB$ and $A$ on $\Omega$, so that (\ref{EH-2}) gives the combination $e^{-\delta} \dot t$. We cannot expect to obtain the two factors separately because of the gauge freedom, which we can use to set $\delta=0$ on $\Omega$.

\medskip

 By (\ref{EH-1}) we have $B\eqq O(a^4/\rho^4)$ and by (\ref{EH-2}) $\nabla_nB\eqq (\rho R)^{-1}O(a^4/\rho^4)$ and, to say that the data is close to Schwarzschild data, we want these to be small. The first term is small if $\rho\gg a$. The second term has dimension (length)$^{-1}$ and will increase without bound as $R$ decreases to zero in the contracting direction. This will only happen inside the horizon. If we restrict $R$ by its value when a marginally-outer-trapped surface forms on $\Omega$ then, from the Friedman equation and with $\rho\gg a$, this happens when 
 \[R^2\rho^2\sim \kappa \mu_0\rho^4,\]
so that we control $\nabla_nB$ on $\Omega$ by controlling $\mu_0$. Now by choice of the location of $\Omega$, at $\rho=\rho_0$, and choice of $\mu_0$ we can choose data close to Schwarzschild. 

\subsubsection{The $k$-Eguchi-Hanson metric}
By this we mean the metric of Pedersen \cite{Pederson}, which can be regarded as the Eguchi-Hanson metric with a cosmological constant ($k$ rather than $\Lambda$, with our conventions), given by
\begin{equation}
\label{LEH-metric}
h_{kEH}=\Delta^{-1}d\rho^2+\frac{\rho^2}{4}(\sigma_1^2+\sigma_2^2)+\frac{\rho^2}{4}\Delta\sigma_3^2,
\end{equation}
where 
$\displaystyle{\Delta=1-\frac{a^4}{\rho^4}-\frac{k}{6}\rho^2}$. This metric is complete for $k<0$ and 
\[a^4=\frac{4}{3k^2}(p-2)^2(p+1), \quad \rho>\left(-\frac{2(p-2)}{k}\right)^{\frac{1}{2}},\]
where $p\geq 3$ in an integer. Then the singularity at $\Delta=0$ is a removable bolt and the level sets of $\rho$ are topologically $S^3/\mathbb{Z}_p$. Since $k$ is related to $a$ for a complete solution, we cannot obtain the previous case from this case by taking $k\rightarrow 0$. However, the matching formulae do formally allow this limit, as we shall see.

Now the Einstein equations for dust source reduce to
\be\label{ev7}\mu R^4=\mu_0,\quad\dot{R}^2+\frac{k}{3}=\frac{\kappa\mu_0}{6R^2}.\ee
We again take the matching surface at constant $\rho$ so that
the first fundamental form on  $\Omega^-$ is 
\[
ds^{2-}|_{\Omega^-}= -\d\t^2+R^2(\tau) \left(\frac{\rho^2}{4}(\sigma_1^2+\sigma_2^2)+\frac{\rho^2}{4}\Delta\sigma_3^2\right).
\]
The equality of the first fundamental forms on $\Omega$ gives
\begin{eqnarray}
\label{LEH-1}
r &\eqq& R\rho  e^{-B},\nonumber\\
e^{-6B} &\eqq& \Delta.
\end{eqnarray}
The normal vector to the matching surface is
\[
n^{-}=\frac{1}{R}\Delta^{\frac{1}{2}}\partial_\rho,
\]
and the associated non-zero components of the second fundamental form on $\Omega^-$ are
\begin{eqnarray}
K_{11}^-=K_{22}^-&\eqq& \frac{\rho R}{4}\Delta^{\frac{1}{2}},\nonumber\\
K_{33}^- &\eqq&   \frac{R}{4}\left(\rho+\frac{a^4}{\rho^3}-\frac{k}{3}\rho^3\right)\Delta^{\frac{1}{2}}.   \nonumber
\end{eqnarray}
The equality of the second fundamental forms gives 
\begin{eqnarray}
\label{LEH-2}
\nabla_{n}B&\eqq&-\frac{\Delta^{-\frac{1}{2}}}{3\rho R}\left(\frac{2a^4}{\rho^4}-\frac{k\rho^2}{6}\right), 
\nonumber\\
Ae^{-\delta} \dot t &\eqq&e^{2B}\left(1-\frac{a^4}{3\rho^4}-\frac{2k\rho^2}{9}\right). 
\end{eqnarray}
We can calculate $A$ as before, to find
\be\label{keh5}
A\eqq  e^{4B}\left(\left(1-\frac{a^4}{3\rho^4}-\frac{2k\rho^2}{9}\right)^2+\frac{k\rho^2}{3}\Delta\right)-\frac{\kappa\mu_0\rho^4e^{-4B}}{6r^2},
\ee
and as before check that this is consistent with $\dot A$ calculated from (\ref{EFE-BCS1}) and (\ref{EFE-BCS2}).

It is not so clear that we may choose data close to Schwarzschild data in this case. We can take $B\eqq 0$, but then the normal derivative is 
$$\nabla_{n}B \eqq  \frac{k\rho}{6R},$$
so that, for this to be small, we would require $R$ to be large on $\Omega$ outside the marginally trapped surface. It is hard to see how to arrange this and so, although the solution in the exterior exists locally, we don't have a good reason to think that it will settle down to Schwarzschild.

%
%
\subsubsection{$k$-Taub-NUT}
We take the Riemannian Taub-NUT metric with a cosmological constant ($k$ rather than $\Lambda$ with our conventions) \cite{bj,akb}
\begin{equation}
\label{LTN-metric}
h_{TN}=\frac{1}{4}\Sigma^{-1}d\rho^2+\frac{1}{4}(\rho^2-L^2)(\sigma_1^2+\sigma_2^2)+L^2\Sigma\sigma_3^2,
\end{equation}
where
\[
\Sigma=\frac{(\rho-L)(1-\frac{k}{12}(\rho-L)(\rho+3L))}{\rho+L},
\]
and use it to contruct the interior:
\[ds^{2-}=-\d\t^2+R^2(\t)h_{TN}.\]
The Einstein equations for a dust source are again (\ref{ev7}).
At the matching surface $\rho=\rho_0$ the first fundamental form is
\[
ds^{2-}|_{\Omega^-}=-d\tau^2+R^2(\tau)\left(\frac{1}{4}(\rho^2-L^2)(\sigma_1^2+\sigma_2^2)+L^2\Sigma\sigma_3^2\right).
\]
From matching the first fundamental forms we get
\begin{eqnarray}
\label{LTN-1}
r & \eqq& R(\rho^2-L^2)^{\frac{1}{2}}e^{-B},\nonumber\\
e^{-6B} & \eqq& \frac{4L^2\Sigma}{\rho^2-L^2}.
\end{eqnarray}
The normal vector to $\Omega^-$ is taken to be
\[
n^{-}=\frac{2}{R}\Sigma^{\frac{1}{2}}\partial_\rho,
\]
and the non-zero components of the second fundamental form in this case are
\begin{eqnarray}
K_{11}^-=K_{22}^- &\eqq& \frac12 R\Sigma^{\frac{1}{2}}\rho, \nonumber\\
K_{33}^- &\eqq& RL^2\Sigma^{\frac{1}{2}}\left(\frac{2\Sigma L}{\rho^2-L^2}-\frac{k (\rho-L)}{6} \right).
\end{eqnarray}
The second matching conditions read
\begin{eqnarray}
\label{LTN-2}
Ae^{-\delta}\dot t &\eqq& \frac{4R}{3r}\Sigma^{\frac{1}{2}}e^{4B}\left(\frac{2L^2(2\rho+L)\Sigma}{\rho^2-L^2}-\frac{k}{6}L^2(\rho-L) \right),\nonumber\\
\nabla_n B& \eqq & \frac{4R}{3r^2}\Sigma^{\frac{1}{2}}e^{4B}\left(\frac{2L^2\Sigma}{\rho+L}+\frac{k}{6}L^2(\rho-L) \right)\\
&\eqq &\frac{1}{3R}\left(\frac{2\Sigma^{1/2}}{\rho+L}+\frac{k(\rho-L)}{6\Sigma^{1/2}}\right).
\end{eqnarray}
We calculate $A$ on $\Omega$ as before and obtain an expression of the form
\[A=c_1(\rho)+\frac{c_2(\rho)}{r^2}\]
and, as before, we can check that this is consistent with $\dot A$ calculated from (\ref{EFE-BCS1}) and (\ref{EFE-BCS2}).

Now note that if $kL^2=-3$ then the metric (\ref{LTN-metric}) is precisely the 4-dimensional hyperbolic metric. In this case, $B$ and $\nabla_nB$ vanish on $\Omega$ whatever the value of $\rho_0$, so that the exterior metric is precisely Schwarzschild: this is a case from section 2 as the interior is now a standard FLRW cosmology. Consequently, if we take $kL^2$ close to $-3$ we expect to get data close to Schwarzschild data. To see that this is the case, set
\[kL^2=-3(1+\epsilon).\]
Then
\[\Sigma = \frac{(\rho^2-L^2)}{4L^2}\left(1+O(\epsilon)\right),\]
so that
\[e^{-6B}\eqq 1+O(\epsilon),   \]
and
\[\nabla_nB\eqq \frac1{LR}O(\epsilon).    \]
Now, clearly the data $(B,\nabla_nB)$ can be chosen as small as desired by choosing large $\rho_0$ and small $\epsilon$.

\bigskip
\bigskip

{\bf Acknowledgements}
We thank CRUP/British Council for Treaty of Windsor grant B-29/08.
FM was supported by CMAT, University of Minho. JN was supported by FCT (Portugal) through program POCI 2010/FEDER and grant POCI/MAT/58549/2004. FM and JN thank the EPSRC and the Oxford Centre for Nonlinear PDE (EP/E035027/1) where this work was initiated, for hospitality.
FM and PT thank Dep. Matem\'atica, Instituto Superior T\'ecnico, for hospitality.

\end{document}